\pdfoutput=1
\documentclass[english,article,prl,reprint,microtype,longbibliography,superscriptaddress]{revtex4-1}
\usepackage[LGR,T1]{fontenc}
\usepackage{color}
\usepackage{babel}
\usepackage{booktabs}
\usepackage{amstext}
\usepackage{graphicx}
\usepackage{subscript}
\usepackage[unicode=true,
 bookmarks=true,bookmarksnumbered=false,bookmarksopen=false,
 breaklinks=false,pdfborder={0 0 0},backref=false,colorlinks=true]
 {hyperref}
\hypersetup{
 pdfauthor={Ambroise van Roekeghem}}

\makeatletter

\DeclareRobustCommand{\greektext}{%
  \fontencoding{LGR}\selectfont\def\encodingdefault{LGR}}
\DeclareRobustCommand{\textgreek}[1]{\leavevmode{\greektext #1}}
\DeclareFontEncoding{LGR}{}{}
\DeclareTextSymbol{\~}{LGR}{126}
\providecommand{\tabularnewline}{\\}

\@ifundefined{date}{}{\date{}}
\date{}

\makeatother

\begin{document}

\title{High throughput thermal conductivity of high temperature solid phases:\\The
case of oxide and fluoride perovskites}

\author{Ambroise van Roekeghem}

\affiliation{CEA, LITEN, 17 Rue des Martyrs, 38054 Grenoble, France}

\email{ambroise.van-roekeghem@polytechnique.edu}

\author{Jes\'{u}s Carrete}

\affiliation{CEA, LITEN, 17 Rue des Martyrs, 38054 Grenoble, France}

\author{Corey Oses}

\affiliation{Center for Materials Genomics, Duke University, Durham, NC 27708,
USA}

\author{Stefano Curtarolo}

\affiliation{Center for Materials Genomics, Duke University, Durham, NC 27708,
USA}

\affiliation{Materials Science, Electrical Engineering, Physics and Chemistry,
Duke University, Durham, NC 27708, USA}

\author{Natalio Mingo}

\affiliation{CEA, LITEN, 17 Rue des Martyrs, 38054 Grenoble, France}

\date{\today}
\begin{abstract}
Using finite-temperature phonon calculations and machine-learning
methods, we calculate the mechanical stability of about 400 semiconducting
oxides and fluorides with cubic perovskite structures at 0\,K, 300\,K
and 1000\,K. We find 92 mechanically stable compounds at high temperatures
-- including 36 not mentioned in the literature so far -- for which
we calculate the thermal conductivity. We demonstrate that the thermal
conductivity is generally smaller in fluorides than in oxides, largely
due to a lower ionic charge, and describe simple structural descriptors
that are correlated with its magnitude. Furthermore, we show that
the thermal conductivities of most cubic perovskites decrease more
slowly than the usual $T^{-1}$ behavior. Within this set, we also
screen for materials exhibiting negative thermal expansion. Finally,
we describe a strategy to accelerate the discovery of mechanically
stable compounds at high temperatures.
\end{abstract}
\maketitle

\section{Introduction}

\label{sec:Introduction}

High throughput ab-initio screening of materials is a new and rapidly
growing discipline \citep{Curtarolo_HT_Nmat}. Amongst the basic properties
of materials, thermal conductivity is a particularly relevant one.
Thermal management is a crucial factor to a vast range of technologies,
including power electronics, CMOS interconnects, thermoelectric energy
conversion, phase change memories, turbine thermal coatings and many
other \citep{Cahill_nanoscale_thermal_transport}. Thus, rapid determination
of thermal conductivity for large pools of compounds is a desirable
goal in itself, which may enable the identification of suitable compounds
for targeted applications. A few recent works have investigated thermal
conductivity in a high throughput fashion \citep{Jesus_PRX_prediction_Half_Heuslers,Seko_HT_low_thermal_conductivity}.
A drawback of these studies is that they were restricted to use the
zero kelvin phonon dispersions. This is often fine when the room temperature
phase is mechanically stable at 0\,K. It however poses a problem
for materials whose room or high temperature phase is not the 0\,K
structure: when dealing with structures exhibiting displacive distortions,
including temperature effects in the phonon spectrum is a crucial
necessity.

Such a phenomenon often happens for perovskites. Indeed, the perovskite
structure can exhibit several distortions from the ideal cubic lattice,
which is often responsible for rich phase diagrams. When the structure
is not stable at low temperatures, a simple computation of the phonon
spectrum using forces obtained from density functional theory and
the finite displacement method yields imaginary eigenvalues. This
prevents us from assessing the mechanical stability of those compounds
at high temperatures or calculating their thermal conductivity. Moreover,
taking into account finite-temperature effects in phonon calculations
is currently a very demanding task, especially for a high-throughput
investigation.

In this study, we are interested in the \textit{high-temperature}
properties of perovskites, notably for thermoelectric applications.
For this reason, we focus on perovskites with the highest symmetry
cubic structure, which are most likely to exist at high temperatures
\citep{Landau_Lifshitz_vol5,Howard_perovskites,Thomas_Muller_perovskites_phase_transitions,Cochran_perovskites,Angel_perovskites_phase_transitions}.
We include the effects of anharmonicity in our ab-initio calculations
of mechanical and thermal properties.

\section{Finite-temperature calculations of mechanical stability and thermal
properties}

\label{sec:Finite-T calculations}

\begin{figure}
\begin{centering}
\includegraphics[scale=0.4]{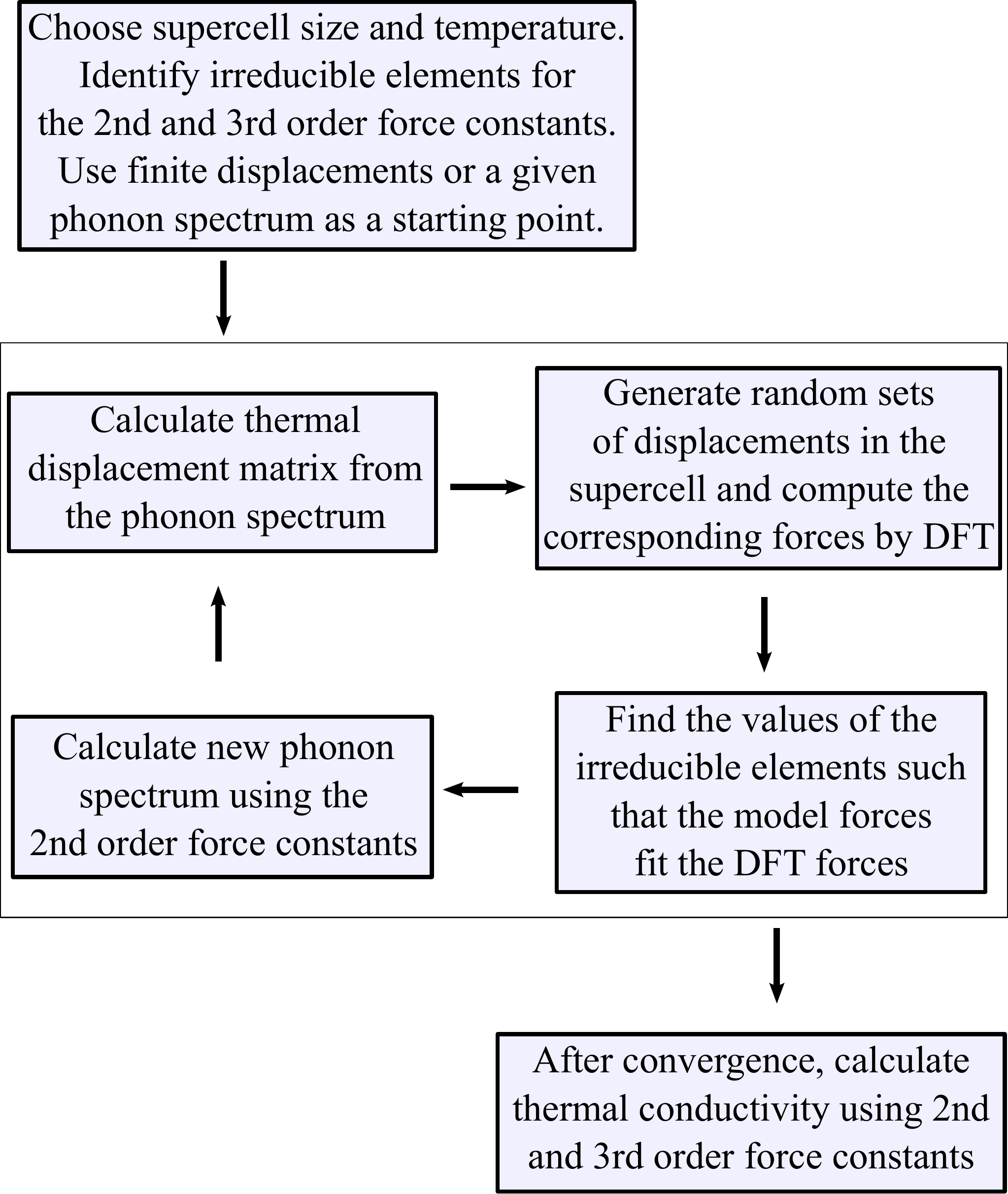}
\par\end{centering}

\caption{Workflow of the method used to calculate the phonon spectrum and thermal
conductivity including finite-temperature anharmonic effects.\label{fig:finite-T-phonon}}
\end{figure}

Recently, several methods have been developed to deal with anharmonic
effects at finite temperatures in solids \citep{Souvatzis_SCAILD,Hellman_TDEP_2011,Hellman_TDEP_2013,Errea_SSCHA,Tadano_SCPH,Ambroise_ScF3}.
In this study, we use the method presented in Ref.\,\citep{Ambroise_ScF3}
to compute the temperature-dependent interatomic force constants,
which uses a regression analysis of forces from density functional
theory coupled with a harmonic model of the quantum canonical ensemble.
This is done in an iterative way to achieve self-consistency of the
phonon spectrum. The workflow is summarized in Fig.\,\ref{fig:finite-T-phonon}.
In the following (in particular Section \ref{sec:PCA-regression}),
it will be referred as ``SCFCS'' -- standing for self-consistent
force constants. As a trade-off between accuracy and throughput, we
choose a 3x3x3 supercell and a cutoff of 5\,Å for the third order
force constants. Special attention is paid to the computation of the
thermal displacement matrix \citep{Ambroise_ScF3}, due to the imaginary
frequencies that can appear during the convergence process, as well
as the size of the supercell that normally prevents us from sampling
the usual soft modes at the corners of the Brillouin zone (see Supplementary
Material). This allows us to assess the stability at 1000\,K of the
391 hypothetical compounds mentioned in Section \ref{sec:Introduction}.
Among this set, we identify 92 mechanically stable compounds, for
which we also check the stability at 300\,K. The phonon spectra of
the stable compounds are provided in the Supplementary Material. Furthermore,
we compute the thermal conductivity using the finite temperature force
constants and the full solution of the Boltzmann transport equation
as implemented in the ShengBTE code \citep{ShengBTE_2014}.

We list the stable compounds and their thermal conductivities in Table
\ref{tab:List-of-perovskites}. Remarkably, this list contains 37
perovskites that have been reported experimentally in the ideal cubic
structure (see References in Table \ref{tab:List-of-perovskites}),
which lends support to our screening method. On the other hand, we
also find that 11 compounds are reported only in a non-perovskite
form. This is not necessarily indicative of mechanical instability,
but instead suggests thermodynamical stability may be an issue for
these compounds, at least near this temperature and pressure. 36 compounds
remain unreported experimentally in the literature to our knowledge.
Thus, by screening only for mechanical stability at high-temperatures,
we reduce the number of potential new perovskites by a factor of 10.
Furthermore, we find that 50 of them are mechanically stable in the
cubic form close to room temperature.

Of the full list of perovskites, only a few measurements of thermal
conductivity are available in the literature. They are displayed in
parentheses in Table \ref{tab:List-of-perovskites} along with their
calculated values. Our method tends to slightly underestimate the
value of the thermal conductivity, due to the compromises we made
to limit the computational cost of the study (see Supplementary Material).
This discrepancy could also be partially related to the electronic
thermal conductivity, which was not substracted in the measurements.
Still, we expect the order of magnitude of the thermal conductivity
and the relative classification of different materials to be consistent.
More importantly, this large dataset allows us to analyze the global
trends driving thermal conductivity. These trends are discussed in
Section\,\ref{sec:Descriptors}.

\begin{table*}
\begin{centering}
\begin{tabular}{ccccclcccccclccccccl}
\toprule 
 & $\kappa_{1000}$ &  & $\kappa_{300}$ &  & Refs. &  &  & $\kappa_{1000}$ &  & $\kappa_{300}$ &  & Refs. &  &  & $\kappa_{1000}$ &  & $\kappa_{300}$ &  & Refs.\tabularnewline
\cmidrule{1-6} \cmidrule{8-13} \cmidrule{15-20} 
\textcolor{blue}{CaSiO$_{3}$} & 4.89 &  &  &  & \citep{Komabayashi_CaSiO3} &  & CdYF$_{3}$ & 1.29 &  & 3.51 &  &  &  & TlOsF$_{3}$ & 0.62 &  & 0.95 &  & \tabularnewline
\textcolor{blue}{RbTaO$_{3}$} & 3.61 &  &  &  & \citep{Lebedev_RbTaO3_RbNbO3} &  & \textcolor{blue}{RbCaF$_{3}$} & 1.15 &  & 2.46 & (3.2) & \citep{Ludekens_RbCaF3_CsCaF3_LiBaF3,Rousseau_RbCaF3,Martin_KZnF3_RbCaF3_KMgF3_KMnF3} &  & InZnF$_{3}$ & 0.61 &  & 1.86 &  & \tabularnewline
\textcolor{blue}{NaTaO$_{3}$} & 3.45 &  &  &  & \citep{Kennedy_NaTaO3} &  & HgInF$_{3}$ & 1.15 &  & 3.85 &  &  &  & \textcolor{blue}{CsCdF$_{3}$} & 0.59 &  & 1.73 &  & \citep{Rousseau_RbCdF3_TlCdF3_CsCdF3}\tabularnewline
\textcolor{red}{CuCF$_{3}$} & 3.32 &  & 8.79 &  & \citep{Zanardi_CuCF3} &  & AlFeF$_{3}$ & 1.14 &  &  &  &  &  & AlMgF$_{3}$ & 0.56 &  &  &  & \tabularnewline
\textcolor{blue}{SrSiO$_{3}$} & 3.23 &  & 10.10 &  & \citep{Xiao_SrSiO3} &  & \textcolor{blue}{PbHfO$_{3}$} & 1.12 &  &  &  & \citep{Kwapulinski_PbHfO3} &  & AuZnF$_{3}$ & 0.53 &  &  &  & \tabularnewline
\textcolor{blue}{NaNbO$_{3}$} & 3.05 &  &  & (\textit{1.5}) & \citep{Shirane_KNbO3_NaNbO3,Mishra_NaNbO3,Tachibana_kappa_perovskites} &  & \textcolor{blue}{AgMgF$_{3}$} & 1.11 &  &  &  & \citep{Portier_AgMgF3_AgMnF3_AgCoF3_AgNiF3_AgCuF3_AgZnF3} &  & InOsF$_{3}$ & 0.52 &  &  &  & \tabularnewline
\textcolor{blue}{BaHfO$_{3}$} & 3.04 &  (4.5) & 8.26 & (10.4) & \citep{Maekawa_BaHfO3_SrHfO3} &  & ZnScF$_{3}$ & 1.10 &  & 3.66 &  &  &  & \textcolor{blue}{RbSrF$_{3}$} & 0.51 &  &  &  & \citep{Pies_Landolt_Bornstein}\tabularnewline
\textcolor{blue}{KNbO$_{3}$} & 2.94 &  &  & (\textit{10}) & \citep{Shirane_KNbO3_NaNbO3,Tachibana_kappa_perovskites} &  & \textcolor{blue}{RbFeF$_{3}$} & 1.09 &  & 4.62 &  & \citep{Kestigian_RbFeF3_CsFeF3} &  & \textcolor{blue}{CsSrF$_{3}$} & 0.50 &  & 1.13 &  & \citep{Pies_Landolt_Bornstein}\tabularnewline
\textcolor{red}{TlTaO$_{3}$} & 2.86 &  &  &  & \citep{Ramadass_TlTaO3_TlNbO3} &  & \textcolor{black}{TlMgF$_{3}$} & 1.06 &  & 3.42 &  & \citep{Arakawa_TlMgF3} &  & BeYF$_{3}$ & 0.48 &  & 2.34 &  & \tabularnewline
\textcolor{blue}{AgTaO$_{3}$} & 2.77 &  &  &  & \citep{Kania_AgTaO3,Pawelczyk_AgTaO3_AgNbO3} &  & \textcolor{blue}{KCaF$_{3}$} & 1.06 &  &  &  & \citep{Demetriou_KCaF3} &  & BeScF$_{3}$ & 0.48 &  & 1.59 &  & \tabularnewline
\textcolor{blue}{KMgF$_{3}$} & 2.74 &  & 8.25 & (10) & \citep{Wood_KMgF3,Martin_KZnF3_RbCaF3_KMgF3_KMnF3} &  & HgScF$_{3}$ & 1.01 &  & 5.42 &  &  &  & \textcolor{blue}{TlCdF$_{3}$} & 0.44 &  &  &  & \citep{Rousseau_RbCdF3_TlCdF3_CsCdF3}\tabularnewline
\textcolor{red}{GaTaO$_{3}$} & 2.63 &  &  &  & \citep{Xu_thesis_GaTaO3,Armiento_perovskites_2011,Castelli_GaTaO3} &  & \textcolor{blue}{CsCaF$_{3}$} & 0.98 &  & 3.03 &  & \citep{Rousseau_KZnF3_phonons} &  & \textcolor{blue}{RbHgF$_{3}$} & 0.43 &  &  &  & \citep{Hoppe_CsHgF3_RbHgF3_KHgF3}\tabularnewline
\textcolor{blue}{BaTiO$_{3}$} & 2.51 &  & 4.99 & (\textit{4-5}) & \citep{Tachibana_kappa_perovskites,Strukov_BaTiO3} &  & AuMgF$_{3}$ & 0.96 &  &  &  & \footnote{AuMgF$_{3}$ was mentioned theoretically in Ref.\,\citep{Uetsuji_AuMgF3}.} &  & PdYF$_{3}$ & 0.43 &  & 0.99 &  & \tabularnewline
\textcolor{blue}{PbTiO$_{3}$} & 2.42 &  &  & (\textit{5}) & \citep{Tachibana_kappa_perovskites} &  & InMgF$_{3}$ & 0.96 &  & 3.53 &  &  &  & AlZnF$_{3}$ & 0.39 &  &  &  & \tabularnewline
\textcolor{blue}{SrTiO$_{3}$} & 2.36 & (4) & 6.44 & (10.5) & \citep{Muta_SrTiO3_kappa,Popuri_SrTiO3_kappa,Yamanaka_SrHfO3} &  & \textcolor{blue}{RbZnF$_{3}$} & 0.91 &  & 2.64 &  & \citep{Daniel_RbZnF3} &  & \textcolor{black}{KHgF$_{3}$} & 0.37 &  &  &  & \citep{Hoppe_CsHgF3_RbHgF3_KHgF3}\tabularnewline
\textcolor{blue}{SrHfO$_{3}$} & 2.20 &  (\textit{2.7}) &  & (\textit{5.2}) & \citep{Kennedy_SrHfO3,Yamanaka_SrHfO3} &  & ZnInF$_{3}$ & 0.88 &  & 1.89 &  &  &  & \textcolor{red}{RbSnF$_{3}$} & 0.37 &  & 0.82 &  & \citep{ThaoTran_NaSnF3_KSnF3_RbSnF3_CsSnF3}\tabularnewline
\textcolor{blue}{BaZrO$_{3}$} & 2.13 &  (2.9) & 5.61 &  (5.2) & \citep{Yamanaka_BaZrO3_kappa} &  & \textcolor{black}{BaSiO$_{3}$} & 0.87 &  &  &  & \citep{Yusa_BaSiO3} &  & ZnBiF$_{3}$ & 0.37 &  & 1.29 &  & \tabularnewline
XeScF$_{3}$ & 1.87 &  & 4.40 &  &  &  & TlCaF$_{3}$ & 0.86 &  &  &  &  &  & \textcolor{blue}{CsHgF$_{3}$} & 0.37 &  & 1.00 &  & \citep{Hoppe_CsHgF3_RbHgF3_KHgF3}\tabularnewline
HgYF$_{3}$ & 1.84 &  & 5.37 &  &  &  & CdScF$_{3}$ & 0.85 &  & 2.37 &  &  &  & \textcolor{red}{KSnF$_{3}$} & 0.35 &  &  &  & \citep{ThaoTran_NaSnF3_KSnF3_RbSnF3_CsSnF3}\tabularnewline
\textcolor{blue}{AgNbO$_{3}$} & 1.79 &  &  &  & \citep{Lukaszewski_AgNbO3,Sciau_AgNbO3} &  & XeBiF$_{3}$ & 0.82 &  & 2.13 &  &  &  & CdBiF$_{3}$ & 0.33 &  & 0.98 &  & \tabularnewline
\textcolor{red}{TlNbO$_{3}$} & 1.75 &  &  &  & \citep{Ramadass_TlTaO3_TlNbO3} &  & \textcolor{blue}{AgZnF$_{3}$} & 0.80 &  &  &  & \citep{Portier_AgMgF3_AgMnF3_AgCoF3_AgNiF3_AgCuF3_AgZnF3} &  & \textcolor{black}{RbPbF$_{3}$} & 0.32 &  &  &  & \citep{Yamane_RbPbF3}\tabularnewline
\textcolor{blue}{KFeF$_{3}$} & 1.72 &  & 6.37 & (3.0) & \citep{Okazaki_KMnF3_KFeF3_KCoF3_KNiF3_KCuF3,Suemune_KFeF3_kappa} &  & PdScF$_{3}$ & 0.79 &  & 1.63 &  &  &  & BeAlF$_{3}$ & 0.30 &  & 1.70 &  & \tabularnewline
SnSiO$_{3}$ & 1.66 &  & 4.22 &  & \citep{Clark_SnSiO3,Armiento_perovskites_2014} &  & \textcolor{blue}{KCdF$_{3}$} & 0.75 &  &  &  & \citep{Hidaka_KCdF3,Hidaka_KCdF3_KRbCdF3} &  & \textcolor{red}{KPbF$_{3}$} & 0.30 &  &  &  & \citep{Hull_KPbF3_RbPbF3_CsPbF3}\tabularnewline
\textcolor{red}{PbSiO$_{3}$} & 1.66 &  & 3.69 &  & \citep{Mackay_PbSiO3,Xiao_PbGeO3} &  & \textcolor{blue}{BaLiF$_{3}$} & 0.73 &  & 2.21 & \footnote{The thermal diffusivity of BaLiF$_{3}$ was measured at 300\,K in
Ref.\,\citep{Duarte_BaLiF3_diffusivity} as $\alpha$=0.037\,cm$^{2}$s$^{-1}$.} & \citep{Mortier_BaLiF3,Duarte_BaLiF3_diffusivity} &  & CsBaF$_{3}$ & 0.29 &  &  &  & \tabularnewline
\textcolor{black}{AuNbO$_{3}$} & 1.56 &  &  &  & \citep{Wu_AuNbO3+} &  & HgBiF$_{3}$ & 0.72 &  & 2.37 &  &  &  & InCdF$_{3}$ & 0.29 &  &  &  & \tabularnewline
\textcolor{red}{CaSeO$_{3}$} & 1.42 &  &  &  & \citep{Wildner_SrSeO3_CaSeO3} &  & ZnAlF$_{3}$ & 0.72 &  & 1.92 &  &  &  & BaCuF$_{3}$ & 0.28 &  &  &  & \tabularnewline
\textcolor{red}{NaBeF$_{3}$} & 1.40 &  & 2.53 &  & \citep{O'Daniel_NaBeF3,Roy_NaBeF3} &  & GaZnF$_{3}$ & 0.69 &  &  &  &  &  & \textcolor{red}{TlSnF$_{3}$} & 0.27 &  & 0.63 &  & \citep{Foulon_TlSnF3}\tabularnewline
\textcolor{blue}{RbMgF$_{3}$} & 1.37 &  & 4.54 &  & \citep{Shafer_RbMgF3} &  & \textcolor{blue}{RbCdF$_{3}$} & 0.68 &  & 1.46 &  & \citep{Rousseau_RbCdF3_TlCdF3_CsCdF3} &  & \textcolor{blue}{TlHgF$_{3}$} & 0.26 &  &  &  & \citep{Hebecker_TlHgF3}\tabularnewline
GaMgF$_{3}$ & 1.34 &  & 2.11 &  &  &  & GaRuF$_{3}$ & 0.67 &  &  &  &  &  & CdSbF$_{3}$ & 0.26 &  &  &  & \tabularnewline
\textcolor{blue}{KZnF$_{3}$} & 1.33 &  & 4.15 &  (5.5) & \citep{Suemune_KZnF3,Martin_KZnF3_RbCaF3_KMgF3_KMnF3} &  & \textcolor{black}{CsZnF$_{3}$} & 0.67 &  & 1.12 &  & \citep{Longo_CsZnF3_CsMnF3_CsFeF3_CsCoF3_CsNiF3_CsMgF3} &  & \textcolor{blue}{TlPbF$_{3}$} & 0.22 &  &  &  & \citep{Buchinskaya_TlPbF3}\tabularnewline
ZnYF$_{3}$ & 1.32 &  & 3.72 &  &  &  & \textcolor{black}{TlZnF$_{3}$} & 0.64 &  & 1.96 &  & \citep{Babel_1967_TlZnF3} &  &  &  &  &  &  & \tabularnewline
\bottomrule
\end{tabular}
\par\end{centering}

\caption{List of cubic perovskites found to be mechanically stable at 1000\,K
and their corresponding computed lattice thermal conductivity (in
W/m/K). We also report the computed lattice thermal conductivity at
300\,K (in W/m/K) when we obtain stability at that temperature. We
highlight in blue the compounds that are experimentally reported in
the ideal cubic perovskite structure, and in red those that are reported
only in non-perovskite structures (references provided in the table).
When no reference is provided, no mention of the compound in this
stoichiometry has been found in the experimental literature. Experimental
measurements of the thermal conductivity are reported in parentheses,
and in italics when the structure is not cubic.\label{tab:List-of-perovskites}}

\end{table*}

We also investigate the (potentially) negative thermal expansion of
these compounds. Indeed, the sign of the coefficient of thermal expansion
$\alpha_{V}$ is the same as the sign of the weighted Gr\"{u}neisen
parameter $\gamma$, following $\alpha_{V}=\frac{\gamma c_{V}\rho}{K_{T}}$,
where $K_{T}$ is the isothermal bulk modulus, $c_{V}$ is the isochoric
heat capacity and $\rho$ is the density \citep{Gruneisen,Ashcroft_Mermin}.
The weighted Gr\"{u}neisen parameter is obtained by summing the contributions
of the mode-dependent Gr\"{u}neisen parameters: $\gamma=\sum\gamma_{i}c_{Vi}/\sum c_{Vi}$.
Finally the mode-dependent parameters are related to the volume variation
of the mode frequency $\omega_{i}$ via $\gamma_{i}=-(V/\omega_{i})(\partial\omega_{i}/\partial V)$.
In our case, we calculate those parameters directly using the second
and third order force constants at a given temperature \citep{Fabian_Si_expansion_Gruneisen,Broido_Si_kappa,Hellman_TDEP_2013}:
\begin{equation}
\gamma_{m}=-\frac{1}{6\omega_{m}^{2}}\sum_{ijk\alpha\beta\gamma}\frac{\epsilon_{mi\alpha}^{*}\epsilon_{mj\beta}}{\sqrt{M_{i}M_{j}}}r_{k}^{\gamma}\Psi_{ijk}^{\alpha\beta\gamma}e^{i\mathbf{q}\cdot\mathbf{r}_{j}}
\end{equation}

This approach has been very successful in predicting the thermal expansion
behavior in the empty perovskite ScF$_{3}$ \citep{Ambroise_ScF3},
which switches from negative to positive around 1100\,K \citep{Greve_ScF3_2010}.
In our list of filled perovskites, we have found only two candidates
with negative thermal expansion around room temperature: TlOsF$_{3}$
and BeYF$_{3}$, and none at 1000\,K. This shows that filling the
perovskite structure is probably detrimental to the negative thermal
expansion.

\begin{figure}
\begin{centering}
\includegraphics[width=8.5cm]{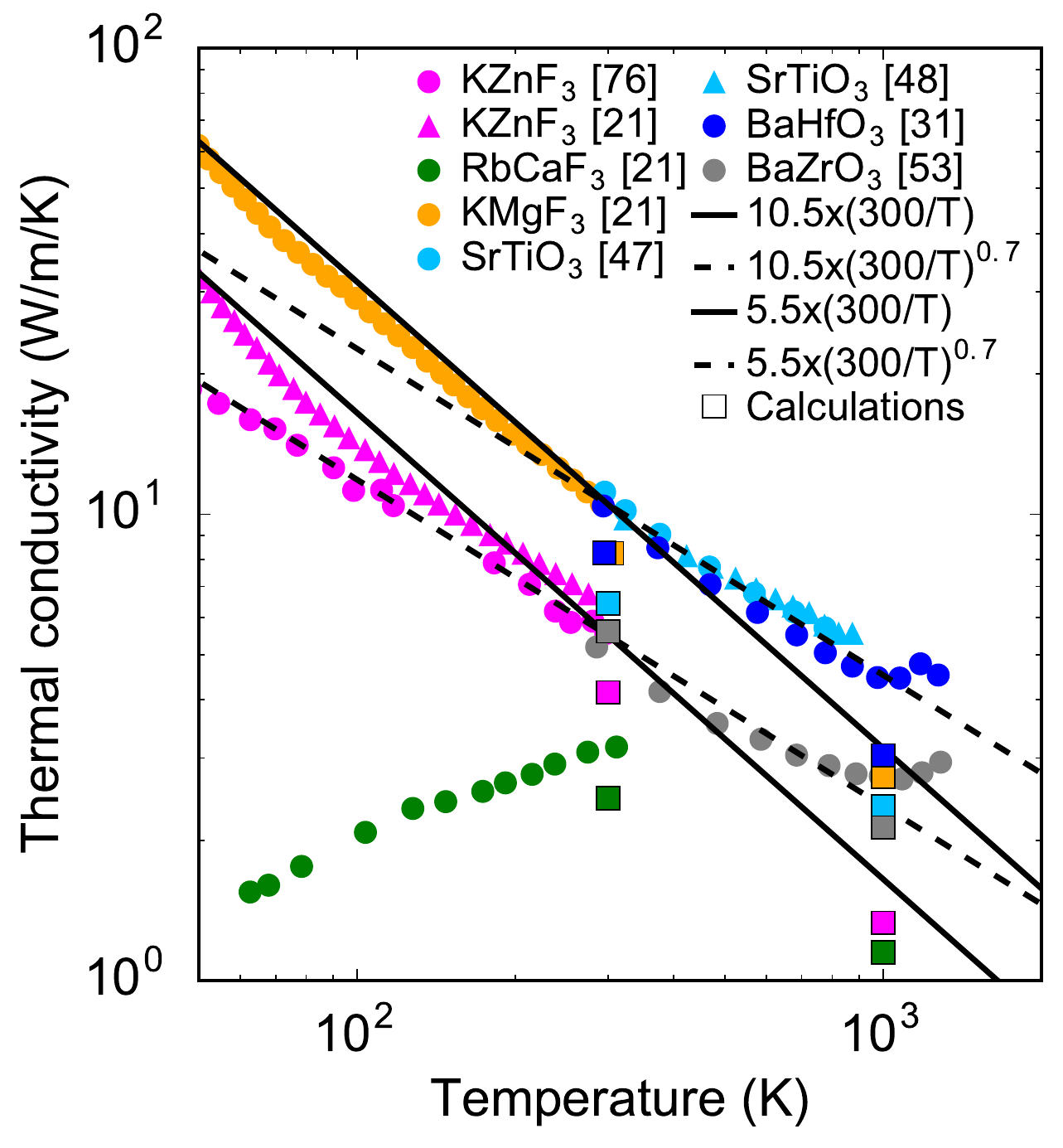}
\par\end{centering}

\caption{A comparison between total thermal conductivities from Refs.\,\citep{Suemune_KZnF3,Martin_KZnF3_RbCaF3_KMgF3_KMnF3,Muta_SrTiO3_kappa,Popuri_SrTiO3_kappa,Maekawa_BaHfO3_SrHfO3,Yamanaka_BaZrO3_kappa},
high-throughput calculations of the lattice thermal conductivity at
300\,K and 1000\,K, and model behaviors in $\kappa\propto T^{-1}$
and $\kappa\propto T^{-0.7}$.\label{fig:Kappa}}
\end{figure}

We also examine the evolution of the thermal conductivity as a function
of temperature, for the compounds that are mechanically stable at
300\,K and 1000\,K. There is substantial evidence that the thermal
conductivity in cubic perovskites generally decreases more slowly
than the model $\kappa\propto T^{-1}$ behavior \citep{Peierls_theorie_cinetique_1929,Roufosse_kappa_high_T}
at high temperatures, in contrast to the thermal conductivity of \textit{e.g.}
Si or Ge that decreases faster than $\kappa\propto T^{-1}$ \citep{Glassbrenner_Si_Ge}.
This happens for instance in SrTiO$_{3}$ \citep{Muta_SrTiO3_kappa,Popuri_SrTiO3_kappa},
KZnF$_{3}$ \citep{Suemune_KZnF3,Martin_KZnF3_RbCaF3_KMgF3_KMnF3},
KMgF$_{3}$ \citep{Martin_KZnF3_RbCaF3_KMgF3_KMnF3}, KFeF$_{3}$
\citep{Suemune_KFeF3_kappa}, RbCaF$_{3}$ \citep{Martin_KZnF3_RbCaF3_KMgF3_KMnF3},
BaHfO$_{3}$ \citep{Maekawa_BaHfO3_SrHfO3}, BaSnO$_{3}$ \citep{Maekawa_BaSnO3}
and BaZrO$_{3}$ \citep{Yamanaka_BaZrO3_kappa}. We also predicted
an anomalous behavior in ScF$_{3}$ using ab-initio calculations,
tracing its origin to the important anharmonicity of the soft modes
\citep{Ambroise_ScF3}. Fig.\,\ref{fig:Kappa} displays several experimentally
measured thermal conductivities from the literature on a logarithmic
scale, along with the results of our high-throughput calculations.
As discussed above, the absolute values of the calculated thermal
conductivities are generally underestimated, but their relative magnitude
and the overall temperature dependence are generally consistent. Although
the behavior of the thermal conductivity $\kappa(T)$ is in general
more complex than a simple power-law behavior, we model the deviation
to the $\kappa\propto T^{-1}$ law by using a parameter $\alpha$
that describes approximately the temperature-dependence of $\kappa$
between 300\,K and 1000\,K as $\kappa\propto T^{-\alpha}$. For
instance, in Fig.\,\ref{fig:Kappa}, KMgF$_{3}$ appears to have
the fastest decreasing thermal conductivity with $\alpha=0.9$ both
from experiment and calculations, while SrTiO$_{3}$ is closer to
$\alpha=0.6$. At present, there are too few experimental measurements
of the thermal conductivities in cubic perovskites to state that the
$\kappa\propto T^{-\alpha}$ behavior with $\alpha<1$ is the general
rule in this family. However, the large number of theoretical predictions
provides a way to assess this trend. Of the 50 compounds that we found
to be mechanically stable at room temperature, we find a mean $\alpha\simeq0.85$,
suggesting that this behavior is likely general and correlated to
structural characteristics of the perovskites.

\section{Accelerating the discovery of stable compounds at high temperature}

\label{sec:PCA-regression}

Through brute-force calculations of the initial list of 391 compounds,
we extracted 92 that are mechanically stable at 1000\,K. However,
this type of calculation is computationally expensive. Thus, it is
desirable for future high-throughput searches of other material classes
to define a strategy for exploring specific parts of the full combinatorial
space. In this section, we propose and test such a strategy based
on an iterative machine-learning scheme using principal component
analysis and regression.

\begin{figure}
\begin{centering}
\includegraphics[width=8.5cm]{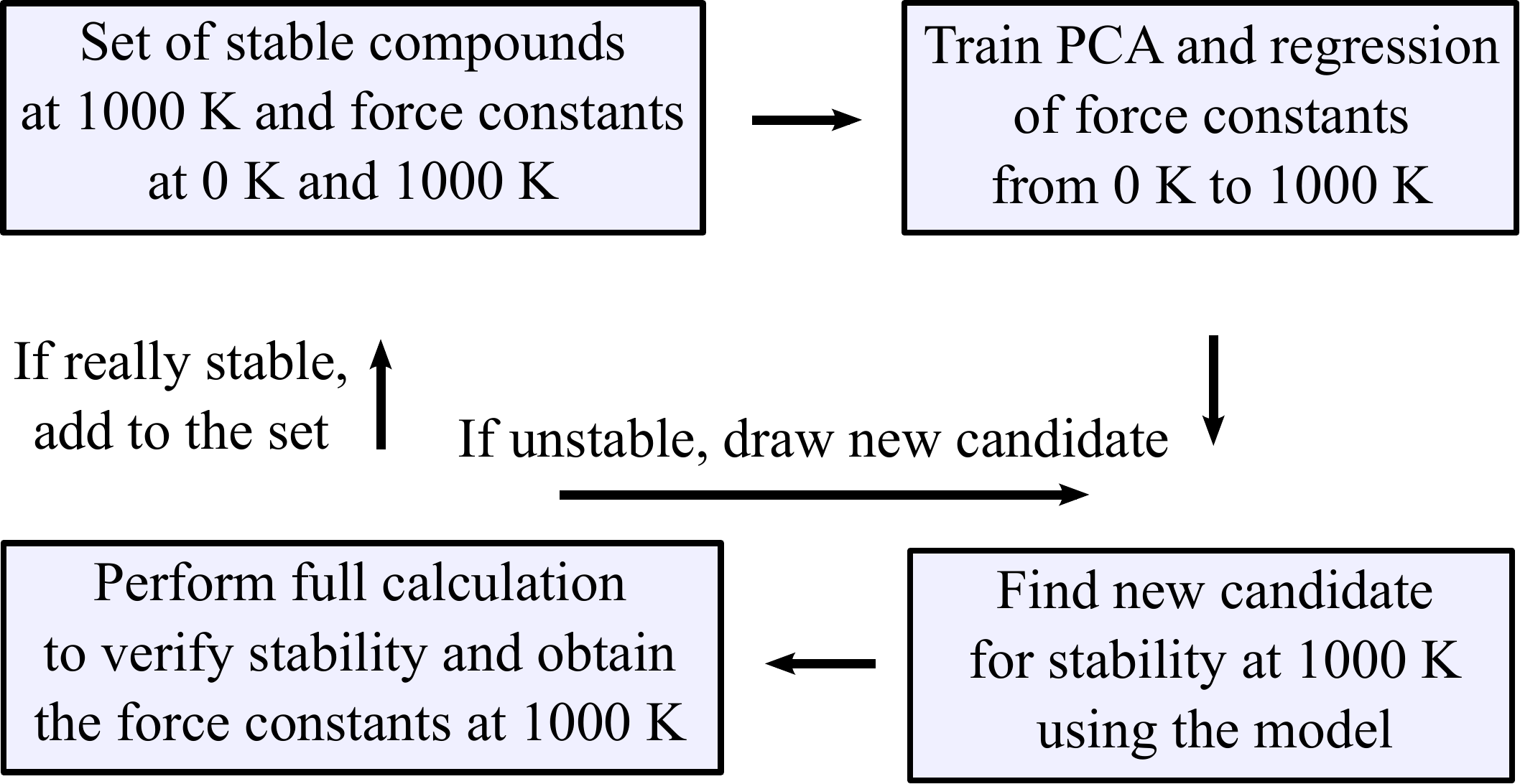}
\par\end{centering}

\caption{Depiction of strategy for exploring the relevant combinatorial space
of compounds that are mechanically stable at high temperature.\label{fig:PCA-regression}}
\end{figure}

We begin by calculating the second order force constants $\Phi_{0\text{\,}K}$
of all compounds using the finite displacement method, which is more
than an order of magnitude faster than finite-temperature calculations.
This gives us a list of 29 perovskites that are mechanically stable
in the cubic phase at 0\,K. Since this is the highest symmetry phase,
they are likely also mechanically stable at high-temperatures \footnote{However, we note that transitions to other structures can take place,
in particular with one of hexagonal symmetry, such as in BaTiO$_{3}$
\citep{Glaister_BaTiO3_hexagonal}, RbZnF$_{3}$\citep{Daniel_RbZnF3}
or RbMgF$_{3}$ \citep{Shafer_RbMgF3}. This phase transition is of
first order, in contrast to displacive transitions that are of second
order.}. We calculate their self-consistent finite-temperature force constants
$\Phi{}_{1000\text{\,}K}^{SCFCS}$ as described in Section \ref{sec:Finite-T calculations}.
This initial set allows us to perform principal component analysis
of the 0\,K force constants so that we obtain a transformation that
retains the 10 most important components. In a second step, we use
regression analysis to find a relation between the principal components
at 0\,K and at 1000\,K. This finally gives us a model that extracts
the principal components of the force constants at 0\,K, interpolate
their values at 1000\,K, and reconstruct the full force constants
matrix at 1000\,K $\Phi_{1000\text{\,}K}^{model}$. We say that this
model has been ``trained'' on the particular set of compounds described
above. Applying it to the previously calculated $\Phi_{0\text{\,}K}$
for all compounds, we can efficiently span the full combinatorial
space to search for new perovskites with a phonon spectrum that is
unstable at 0\,K but stable at 1000\,K. For materials determined
mechanically stable with $\Phi_{1000\text{\,}K}^{model}$, we calculate
$\Phi_{1000\text{\,}K}^{SCFCS}$. If the mechanical stability is confirmed,
we add the new compound to the initial set and subsequently train
the model again with the enlarged set. When no new compounds with
confirmed mechanical stability at high temperatures are found, we
stop the search. This process is summarized in Figure \ref{fig:PCA-regression}.
Following this strategy, we find 79 perovskites that are stable according
to the model, 68 of which are confirmed to be stable by the full calculation.
This means that we have reduced the total number of finite-temperature
calculations by a factor of 5, and that we have retrieved mechanically
stable compounds with a precision of 86\% and a recall of 74\% \footnote{Precision is defined as the fraction of true positives in all positives
reported by the model and recall as the fraction of true positives
found using the model with respect to all true positives.}. It allows us to obtain approximate phonon spectra for unstable compounds,
which is not possible with our finite-temperature calculations scheme
(see Supplementary Material). It also allows us to find compounds
that had not been identified as mechanically stable by the first exhaustive
search due to failures in the workflow. Considering the generality
of the approach, we expect this method to be applicable to other families
of compounds as well. Most importantly, it reduces the computational
requirements, particularly if the total combinatorial space is much
larger than the space of interest.

\section{Simple descriptors of the thermal conductivity}

\label{sec:Descriptors}

\begin{figure}
\begin{centering}
\includegraphics[width=8.5cm]{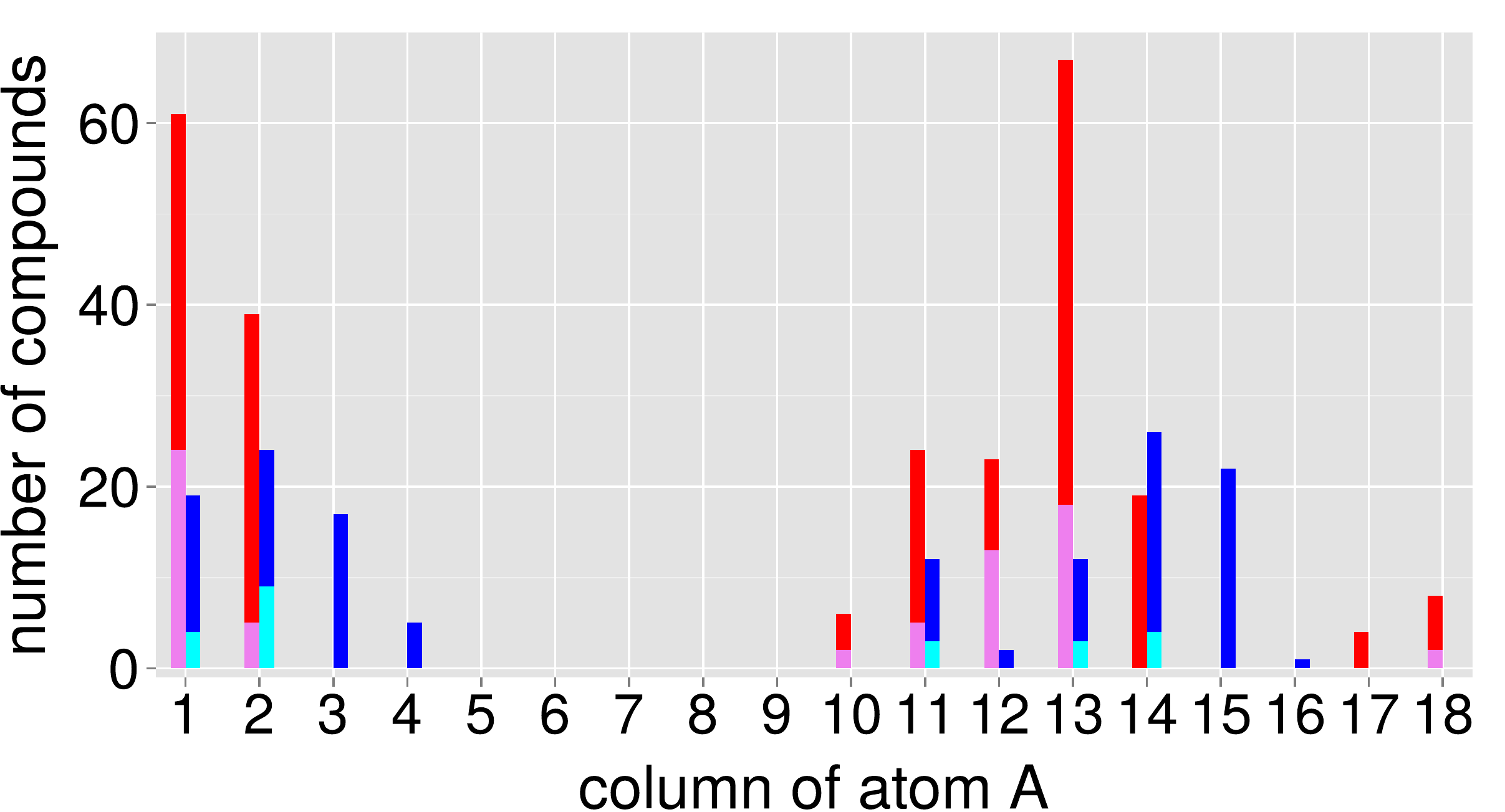}
\par\end{centering}

\begin{centering}
\includegraphics[width=8.5cm]{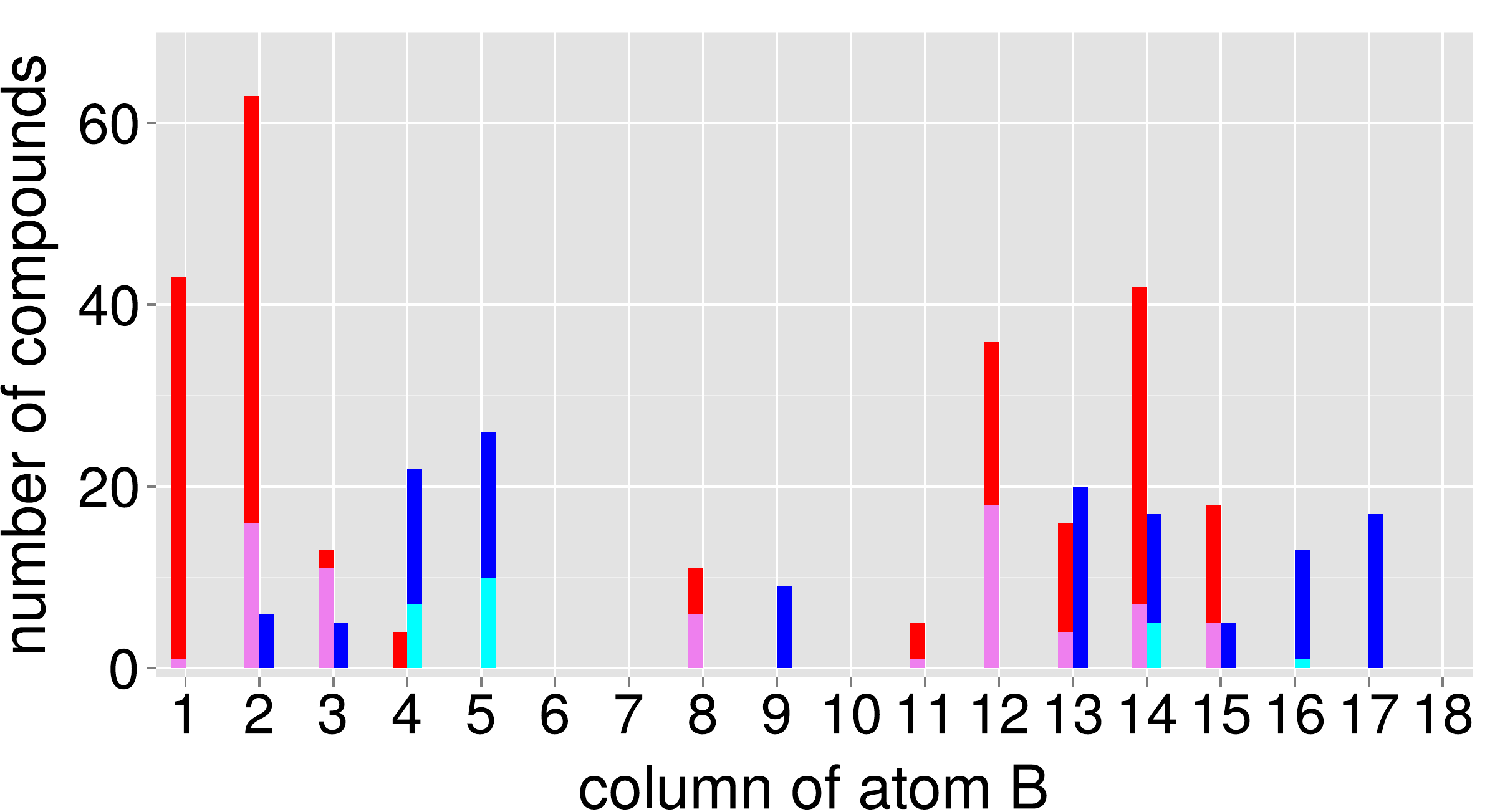}
\par\end{centering}

\caption{Column number of the element at site \textit{A} (top) and \textit{B}
(bottom) of the perovskite \textit{ABX}$_{3}$ in the initial list
of fluorides (red) and oxides (blue) paramagnetic semiconductors and
after screening for mechanical stability (violet and cyan, respectively).\label{fig:Columns}}
\end{figure}

We now focus on the analysis of the thermal conductivity data provided
in Table\,\ref{tab:List-of-perovskites}. We note that this set contains
about two times more fluorides than oxides. This was already the case
after the first screening in which we kept only the semiconductors,
and it can be explained by the strong electronegativity of fluorine,
which generally forms ionic solids with the alkali and alkaline earth
metals easily, as well as with elements from groups 12, 13 and 14.
This is shown on Fig.\,\ref{fig:Columns}, in which we display histograms
of the columns of elements at sites \textit{A} and \textit{B} of the
perovskite in our initial list of paramagnetic semiconductors and
after screening for mechanical stability.

\begin{figure}
\begin{centering}
\includegraphics[width=8.5cm]{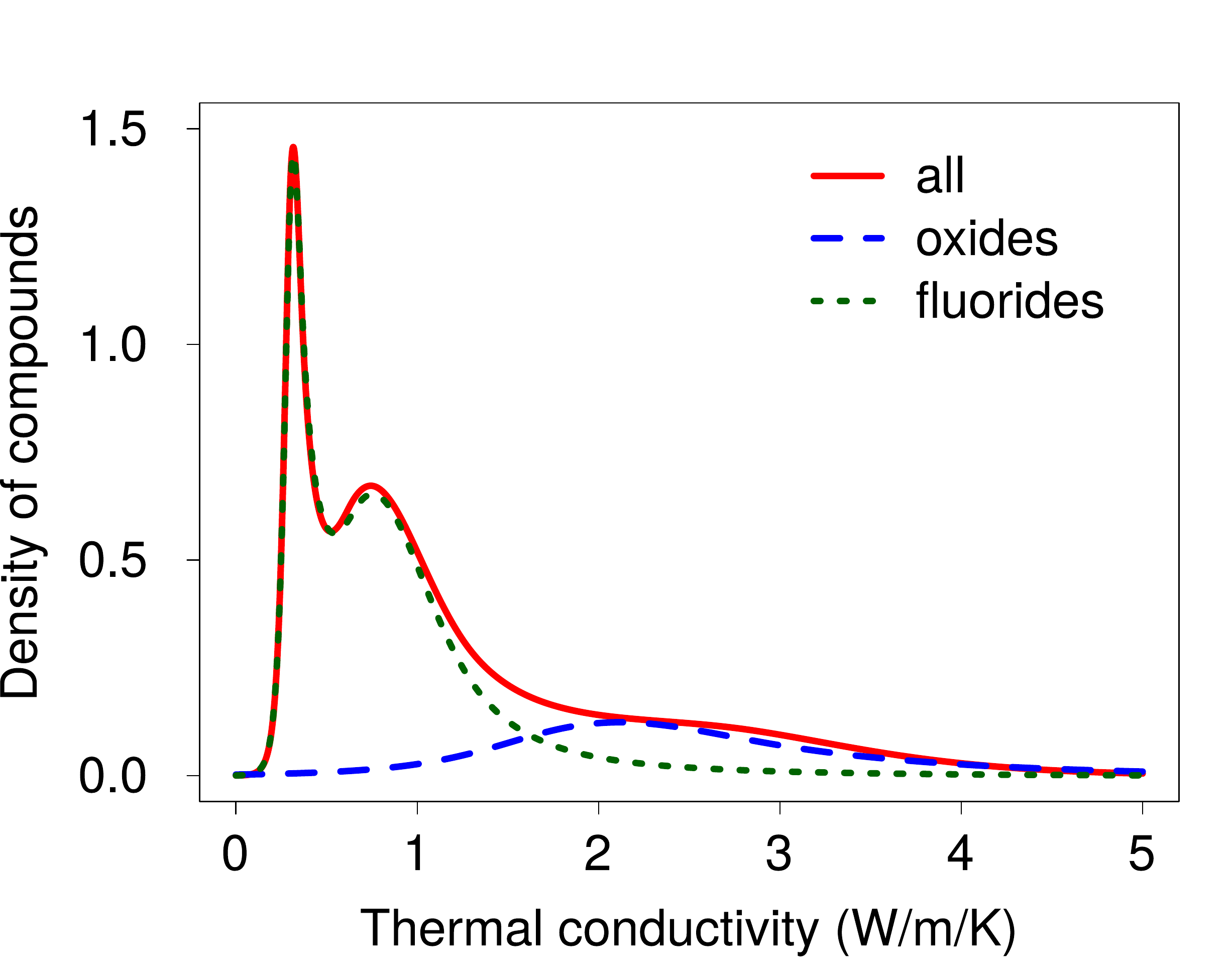}
\par\end{centering}

\caption{Distribution of compounds as a function of the lattice thermal conductivity
at 1000\,K. The red curve corresponds to the distribution for all
mechanically stable compounds. The blue curve corresponds to the distribution
for fluorides only. The green curve corresponds to the distribution
for oxides only.\label{fig:Fluorides_vs_oxides}}
\end{figure}

We can also see that the oxides tend to display a higher thermal conductivity
than the fluorides, as shown on the density plot of Fig.\,\ref{fig:Fluorides_vs_oxides}.
This is once again due to the charge of the fluorine ion, which is
half that of the oxygen ion. In a model of a purely ionic solid, this
would cause the interatomic forces created by electrostatic interactions
to be divided by two in fluorides as compared to oxides. This is roughly
what we observe in our calculations of the second order force constants.
It translates into smaller phonon frequencies and mean group velocities
in fluorides as compared to oxides. Fluorides also have smaller heat
capacities, due to their larger lattice parameters (see Supplementary
material). Those two factors mainly drive the important discrepancy
of the thermal conductivity between fluorides and oxides. Following
the same reasoning, it means that halide perovskites in general should
have a very low thermal conductivity.

\begin{figure*}
\begin{centering}
\includegraphics[width=8.5cm]{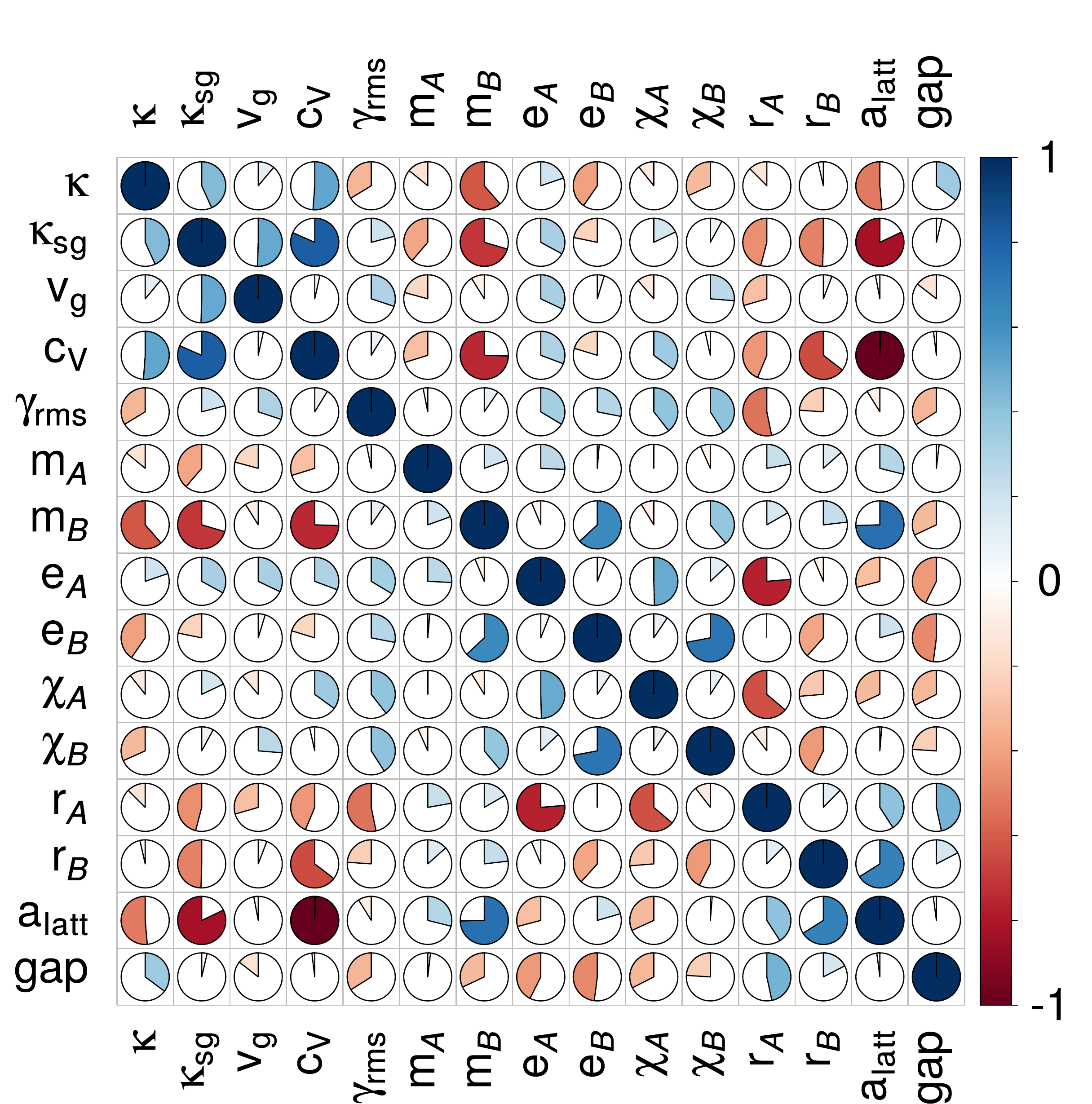}\includegraphics[width=8.5cm]{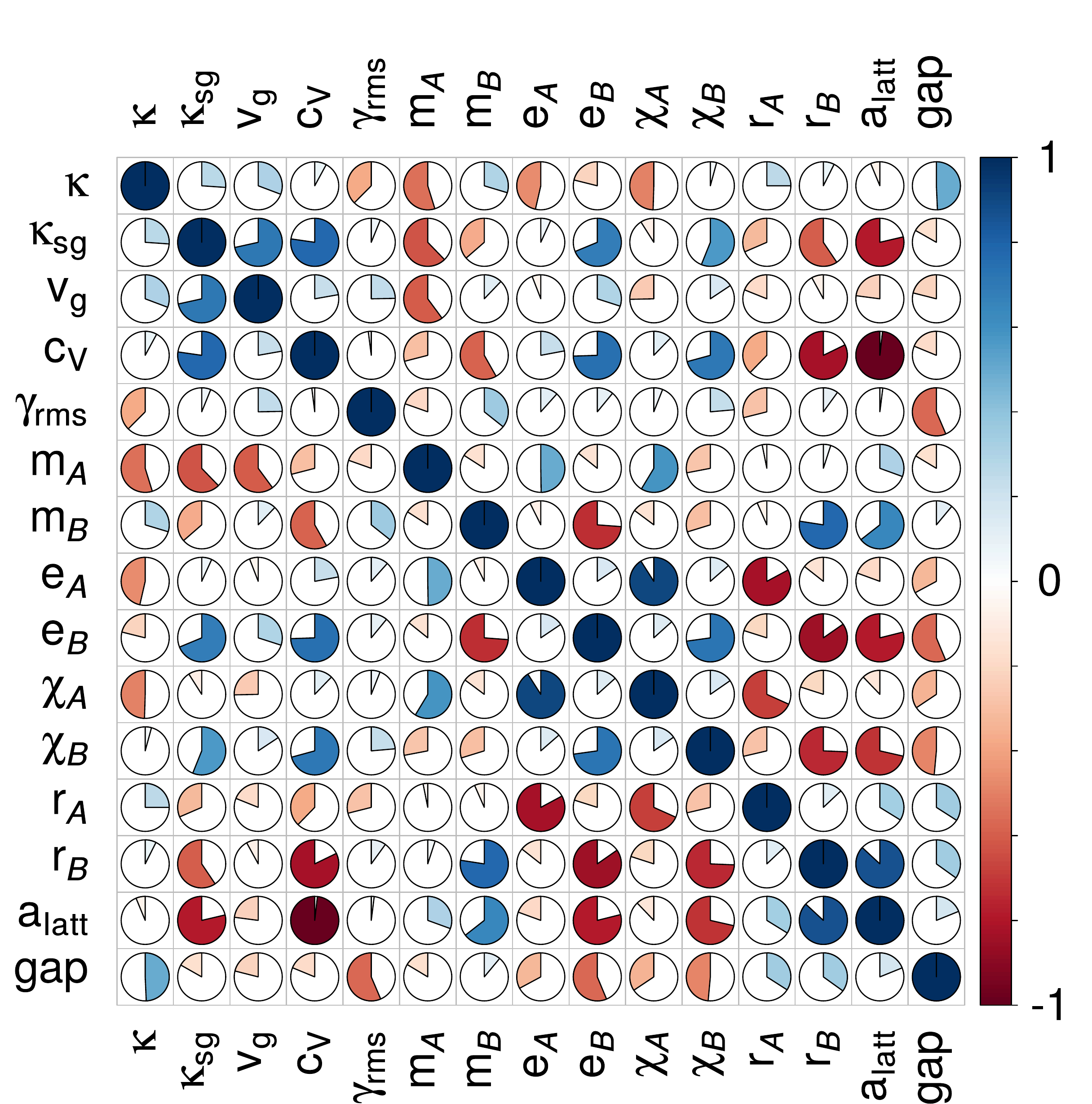}
\par\end{centering}

\caption{Correlograms between the thermal conductivity $\kappa$, the thermal
conductivity in the small grain limit $\kappa_{\text{sg}}$, the mean
phonon group velocity v$_{\text{g}}$, the heat capacity c$_{\text{V}}$,
the root mean square Gr\"uneisen parameter $\gamma_{\text{rms}}$,
the masses m\protect\textsubscript{\textit{A}} and m\protect\textsubscript{\textit{B}}
of atoms at sites \textit{A} and \textit{B} of the perovskite \textit{ABX}$_{3}$,
their electronegativity e\protect\textsubscript{\textit{A}}, e\protect\textsubscript{\textit{B}},
their Pettifor scale \textgreek{q}\protect\textsubscript{\textit{A}},
\textgreek{q}\protect\textsubscript{\textit{B}}, their ionic radius
r\protect\textsubscript{\textit{A}}, r\protect\textsubscript{\textit{B}},
the lattice parameter of the compound a\protect\textsubscript{latt}
and its electronic gap, for mechanically stable fluorides (left) and
oxides (right) at 1000\,K.\label{fig:Correlograms}}
\end{figure*}

Finally, we analyze the correlations between the thermal conductivity
and different simple structural descriptors. Fig.\,\ref{fig:Correlograms}
displays the correlograms for fluorides and oxides between the following
variables: the thermal conductivity $\kappa$, the thermal conductivity
in the small grain limit $\kappa_{\text{sg}}$ \citep{Carrete_Nanograined,Jesus_PRX_prediction_Half_Heuslers},
the mean phonon group velocity v$_{\text{g}}$, the heat capacity
c$_{\text{V}}$, the root mean square Gr\"uneisen parameter $\gamma_{\text{rms}}$
\citep{Bjerg_rms_Gruneisen,Madsen_rms_Gruneisen}, the masses of atoms
at sites \textit{A} and \textit{B} of the perovskite \textit{ABX}$_{3}$,
their electronegativity, their Pettifor number \citep{Pettifor_scale},
their ionic radius, the lattice parameter of the compound and its
electronic gap. Remarkably, sites \textit{A} and \textit{B} play very
different roles in fluorides and oxides. In particular, the thermal
conductivity of fluorides is mostly influenced by substitutions of
the atom inside the fluorine octahedron (site \textit{B}), while the
interstitial atom at site \textit{A} has a negligible impact. The
opposite is true for the oxides. This means that when searching for
new compounds with a low lattice thermal conductivity, substitutions
at the \textit{A} site of fluorides can be performed to optimize cost
or other considerations without impacting thermal transport. It is
also interesting to note that the gap is largely correlated with the
electronegativity of atom \textit{B}, suggesting the first electronic
excitations likely involve electron transfer from the anion to the
\textit{B} atom.

Common to both fluorides and oxides, the lattice parameter is mostly
correlated with the ionic radius of atom \textit{B} rather than atom
\textit{A}. Interestingly, the lattice parameter is larger for fluorides,
although the ionic radius of fluorine is smaller than for oxygen.
This is presumably due to partially covalent bonding in oxides (see
e.g. Ref.\,\onlinecite{Kolezynski_BaTiO3}). In contrast, fluorides
are more ionic: the mean degree of ionicity of the \textit{X-B} bond
calculated from Pauling's electronegativities \citep{Pauling_electronegativity}
$e_{X}$ and $e_{B}$ as $I{}_{XB}=100\left(1-e^{\left(e_{X}-e_{B}\right)/4}\right)$
yields a value of 56\% for oxides versus 74\% for fluorides. Ionicity
is also reflected by the band structure, as can be seen from the weak
dispersion and hybridization of the F-2$p$ bands \footnote{See for instance the band structure of SrTiO$_{3}$ \citep{van_Benthem_SrTiO3}
compared to the one of KCaF$_{3}$ \citep{Ghebouli_KCaF3}. In those
two compounds, the degree of ionicity of the \textit{X}-\textit{B}
bond calculated from Pauling's electronegativity is 59\% and 89\%,
respectively.}. This may explain why the role of atoms at site \textit{A} and \textit{B}
is so different between the two types of perovskites. We think that
the more ionic character combined to the small nominal charge in fluorides
makes the octahedron cage enclosing the atom \textit{B} less rigid,
such that the influence of the atom \textit{B} on the thermal conductivity
becomes more significant. 

\label{sec:Unusual_behavior}

\section{Conclusion}

Employing finite-temperature \textit{ab-initio} calculations of force
constants in combination with machine learning techniques, we have
assessed the mechanical stability and thermal conductivity of hundreds
of oxides and fluorides with cubic perovskite structures at high temperatures.
We have shown that the thermal conductivities of fluorides are generally
much smaller than those of oxides, and we found new potentially stable
perovskite compounds. We have also shown that the thermal conductivity
of cubic perovskites generally decreases more slowly than the inverse
of temperature. Finally, we provide simple ways of tuning the thermal
properties of oxides and fluorides by contrasting the effects of substitutions
at the \textit{A} and \textit{B} sites. We hope that this work will
trigger further interest in halide perovskites for applications that
require a low thermal conductivity.
\begin{acknowledgments}
This work is partially supported by the French ``Carnot'' project
SIEVE. C. Oses acknowledges support from the National Science Foundation
Graduate Research Fellowship under Grant No. DGF1106401. We also acknowledge
the CRAY corporation for computational support.

\end{acknowledgments}

\end{document}